# Rules essential to water molecular undercoordination

11.7 k words, 13 figures, 115 refs

Chang Q Sun (孙长庆)[1,*]

**Abstract**

A sequential of concepts developed in last decade has enabled a resolution to multiple anomalies of water ice and its low-dimensionality, particularly. Developed concepts include the coupled hydrogen bond (O:H−O) oscillator pair, segmental specific heat, three-body coupling potentials, quasisolidity, and supersolidity. Resolved anomalies include ice buoyancy, ice slipperiness, water skin toughness, supercooling and superheating at the nanoscale, etc. Evidence shows consistently that molecular undercoordination shortens the H−O bond and stiffens its phonon while undercoordination does the O:H nonbond contrastingly associated with strong lone pair ":" polarization, which endows the low-dimensional water ice with supersolidity. The supersolid phase is hydrophobic, less dense, viscoelastic, thermally more diffusive and stable, having longer electron and phonon lifetime. The equal number of lone pairs and protons reserves the configuration and orientation of the coupled O:H−O bonds and restricts molecular rotation and proton hopping, which entitles water the simplest, ordered, tetrahedrally-coordinated, fluctuating molecular crystal covered with a supersolid skin. The O:H−O segmental cooperativity and specific-heat disparity form the soul dictating the extraordinary adaptivity, reactivity, recoverability, sensitivity of water ice when subjecting to physical perturbation. It is recommended that the premise of "hydrogen bonding and electronic dynamics" would deepen the insight into the core physics and chemistry of water ice.

---

[1] School of EEE, Nanyang Technological University, Singapore 639798; EBEAM, Yangtze Normal University, Chongqing 408100, China (ecqsun@ntu.edu.sg)

Contents

## 1. Wonders of water structure and its molecular undercoordination

*− Coordination bonding and electronic dynamics govern the performance of substance* [1]

Laid the platform for defect physics, surface chemistry, nanoscience and nanotechnology, atomic or molecular undercoordination amazed the mysterious water ice even further [2-4]. For instances, the skin of water is toughest, and ice is naturally most slippery of ever known [5-9]. The extent of slipperiness and toughness increases with the curvature of the surface, making a nanodroplet and a nanobubble more chemically reactive but mechanically and thermally endurable [10-13]. A water droplet bounces rounds before it disappears when falls on water, showing both skins elastic and hydrophobic. Water droplets encapsulated in the hydrophobic pores and ultrathin water films deposited on the hydrophobic surfaces of graphite, silica, protein, and selected metals behave like ice at the ambient temperature [14-18]. Kept a 0.6 nm thick air-gap between its skin and the wall, water droplet travels through a microchannel at a speed much faster than what the classical fluid theory could expect [19]. The nanometer-sized droplet melts at a temperature of some 50 °K higher than the bulk melting point, $T_m$ = 273 °K [20]; contrastingly, a droplet of 1.4 nm across freezes at 205 °K [21], compared to the bulk freezing point, $T_N$ = 258 °K [22]. Table 1 lists typical anomalies pertained to the skin-and-nanoscale water ice, and the hydrogen bonding and electronic characteristics at the skin.

The low-dimensional water ice is ubiquitously important to the quality and sustainability of human life [8]. Considerable effort has been made to the understanding of anomalies of water and ice, at the molecular and nanoscale level, particularly, since 1859 when Faraday and Thomson [23, 24] firstly noted that a liquid-like layer not only makes ice slippery but also welds two blocks of ice – known as ice regelation [25, 26]. A verity of experimental techniques has been used to explore the performance of molecules, protons, and electrons in the spatial-temporal-energetic domains, such as low-energy electron diffraction (LEED) [27], neutron [21] and proton [28] diffractions, atomic force microscopy (AFM) [20] and scanning tunneling microscopy/spectroscopy (STM/S) [29]. Photoelectron spectroscopy of X-ray (XPS) [30, 31] and ultra-violet (UPS) [32] excitation and phonon spectroscopy of Raman scattering [22], infrared transmission [33], nuclear magnetic resonance (NMR) [34], and sum frequency generation (SFG) [35, 36], have contributed immensely to the advancement of this field.

Theoretically, the classical continuum thermodynamics deals with water and ice as a collection of gaseous-like neutral particles to examine the response of the entire body to stimuli, which succeeds in formulating the liquid-vapor phase transition in terms of enthalpy, entropy, and Gibbs free energy [37, 38]. Molecular dynamics (MD) [39-43] treats the flexible or rigid, polarizable or non-polarizable, individual molecular dipole as the primary unit of structure. Combining MD computations and ultrafast phonon spectroscopy reveals spatial-temporal performance of molecules with derived information of phonon relaxation or the molecular residing time in a specific coordination site, and the manner and mobility of mass transport. An interlay of STM/S and MD simulations of the proton quantum effect has enabled visualization of the two-dimensional ice formation [44] and the concerted tunneling of protons within a water cluster with quantification of the impact of zero-point motion on the strength of single hydrogen bond at a water/solid interface [29]. Density functional theory (DFT) calculations resolve the perturbation derived relaxation of the O:H−O segmental length and vibrational frequencies [22, 45]. The perturbation refers to electrostatic polarization, mechanical compression, molecular undercoordination, thermal excitation, charge injection by aqueous solvation, etc.

This presentation shows the essentiality of a sequential rules and concepts developed in last decade to resolving anomalies of water ice and its low-dimensionality. Focus is on the "local bonding and electronic dynamics" towards deeper understanding of the core physics and chemistry of water ice. Evidence shows consistently that molecular undercoordination driven O:H–O cooperative relaxation, nonbonding electron polarization, and specific-heat dispersion resolve a common supersolid skin that is hydrophobic, less dense, reactive, viscoelastic, and thermally more stable. It is also uncovered that the supersolid skin is thermally more diffusive with a lower specific heat. The supersolidity reconciles the anomalies such as ice floating, ice slipperiness, superfluidity and hydrophobicity of low-dimensional water and ice.

Table 1 Typical examples for the molecular undercoordination resolved anomalies of low-dimensional water ice.

| | | Consequence of molecular undercoordination | Bonding and electronic origin |
|---|---|---|---|
| Thermo-mechanical anomalies | 1 | **Water skin toughness** [46]. The surface stress is 72.75 mJ/m$^2$ compared with 26.6 mJ/m$^2$ for CCl$_4$ solution at 293 °K. Surface stress drops linearly with the rise of temperature [47]. | Supersolidity of higher melting point $T_m$ and lower freezing temperature $T_N$ and polarization. |
| | 2 | **Slipperiness of ice** [9, 48]. The slipperiness of wet surfaces is most for hydrophilic/hydrophobic contact but least for hydrophilic/hydrophilic interaction [49]. Ice on ice has a higher friction coefficient. | O:H nonbond high elastic adaptivity and surface dipolar repulsivity. |
| | 3 | **Hydrophobicity and elasticity** [50, 51]. Water | Skin elastic repulsive |
| | | droplet dances rounds before merging into the bulk when falling on to liquid water. | supersolidity. |
| | 4 | **Ice skin premelting and nanorheology** [6, 52]. The skin is viscoelastic over a wide span of temperature. with a viscosity up to two orders of magnitude larger than pristine water. | Gel-like, viscoelastic supersolid phase. |
| | 5 | **Skin low mass density** [53]. The skin mass density is confirmed 0.75 g/cm$^3$ opposing to classical thermodynamics prediction, XRD revealed 5.9% skin O—O elongation with respect to 2.8 Å length or 15.6% density loss at 298 °K. In contrast, the skin O—O for liquid methanol contracts by 4.6% associated with a 15% density gain [54]. | O:H expands more than H−O contraction. |
| | 6 | **Thermal diffusivity and specific heat** [55]. High thermal diffusivity and low specific heat ensure heat outward flow and high surface temperature in the thermal transport of warm water to cold drain - Mpemba effect [56]. | Skin density-loss dominance of diffusivity and Debye temperature offset by phonon frequency relaxation. |
| | 7 | **Thermal stability** [17, 57]. Raman H–O phonon skin-component at 3450 cm$^{-1}$ is less sensitive to temperature than its bulk at 3200 cm$^{-1}$. Skin and bulk components undergo thermal contraction yet the dangling H–O bond undergoes thermal expansion. | Heating can hardly deform further the undercoordination deformed H−O bond. |
| | 8 | **Nanobubble durability and reactivity** [58-61]. Nanobubble is mechanically and thermally endurable and chemically more reactive. | Skin supersolidity ensures mechanical and thermal stability; polarization raises the reactivity. |
| | 9 | **Supercooling and superheating** [62, 51]. Water nanodroplet or bubbles undergo superheating at melting and supercooling at freezing and evaporating, whose extent is droplet size dependence. A 1.2 nm sized droplet freezes at temperature below 172 °K and the monolayer skin melts at 320 °K. | QS boundary dispersion by O:H−O relaxation through Einstein's relation: $\Delta\Theta_x \propto \Delta\omega_x$, raising the $T_m$ and lowering the $T_V$ and $T_V$. |
| Electron phonon characteristics | 10 | **Electron entrapment and polarization** [63, 64, 31]. O 1$s$ core-level shifts from the bulk value of 536 to 538 for the skin and to 540 eV for the gaseous state. The hydrated nonbonding electron shifts its bound energy from the bulk value of 2.4 for the interior and 1.2 eV for the skin to the limit of 0.4 eV when the $(H_2O)_N^-$ cluster size is reduced to N = 5. | H−O bond contraction deepens the local potential well, which entraps the core levels; densely entrapped core electrons polarizes the nonbonding electrons. |
| | 11 | **Phonon stiffness** [65, 53]. Skins of 298 °K water and (253-258) °K ice share an identical H–O phonon frequency of 3450 cm$^{-1}$, in contrast to the bulk values of 3200(water) and 3150(ice) cm$^{-1}$ and 3650 cm$^{-1}$ for the $H_2O$ monomer in gaseous phase. H−O dangling bond frequency of 3610 cm$^{-1}$. H−O phonon frequency increases linearly with the inverse of cluster size [4]. | Phonon frequency shift is proportional to the square-root of segmental cohesive energy and inversely to the segmental length. |
| | 12 | **Refractive index** [66]. The refractive index of water (measured at $\lambda$ = 589.2 nm) skin is higher than it is in the bulk. | Polarization dominance of dielectric permittivity. |
| | 13 | **Lifetime of skin H−O phonon** [33] **and hydrating electrons** [67, 68, 32, 69, 64]. Skin hydrated electrons and stiffened phonons have longer lifetime or slower relaxation dynamics. | Skin polarization and boundary wave reflection; quasi standing wave formation. |
| Length-energy transition | 14 | **Bond length** [63, 54]. H–O bond contracts from 1.00 to 0.95 Å and the O:H from 1.70 to 1.95 Å; H−O dangling bond length of 0.9 Å. Skin $d_{OO}$ was measured as 2.965 Å compared to the bulk water of 2.70 Å [54]. | |
| | 15 | **Bond energy** [70, 63]. The O:H–O segmental energies transit from (0.2, 4.0) to (0.1, 4.6) eV when moving from the bulk to the skin in comparison to the least-coordinated gaseous H–O bond energy of 5.10 eV. Gaseous H–O dissociation requires 121.6 nm laser beam irradiation [70]. | |

2. Rule 1: O:H−O bond cooperativity versus crystallinity

As the basic functional and interaction elements of molecular crystals, the equally numbered electron lone pairs ":" and dangling protons ($H^+$ simplified as H) entitle water the simplest, ordered, tetrahedrally-coordinated, uniform yet fluctuating structure among known molecular crystals. Represented using a $H_2O:4H_2O$ tetrahedral motif with one $H_2O$ molecular in the center and four on the apical sites, water reserves its O:H−O configuration and orientation over broad pressure and temperature ranges, from 5 to 2000 °K and from $10^{-11}$ to $10^{12}$ Pa [71, 72]. $H_2O$ molecular rotation is subject to restriction. Rotating a $H_2O$ molecule around its $C_{3v}$-axis by 120° causes long-range disorder of its two-dimensional hexagonal latticed ice [73] because of the presence of repulsive H↔H and O:⇔:O interactions [74]. $H^+$ transitional tunneling or hopping is also restricted from one asymmetrical site to the other between adjacent oxygen atoms because dissociating a H−O bond of 5.1 eV energy in the vapor phase requires a 121.6 nm laser radiation [70].

In placing the molecular wise, the O:H−O bond, as the basic functional and structural unit, features the performance of electrons, bonding, and molecules in the energetic-spatial-temporal domains. The strong repulsion between lone pairs of adjacent oxygen anions endows the O:H−O an asymmetrical, short-range, and coupled oscillator pair that integrates the intermolecular O:H nonbond and the intramolecular H−O polar-covalent bond interactions. The O:H−O bond responds to a physical perturbation cooperatively in relaxing its segmental length and energy and electronic dynamics [75].

Fig 1a illustrates manners of O:H−O bond segmental length relaxation under perturbation. Both $O^{2-}$ anions dislocate in the same direction along the O:H−O by different amounts $\Delta d_x$ with respect to the coordination origin $H^+$. Subscript x = L and H represent for the O:H and the H−O, respectively. The O—O distance varies through one segment contraction and the other elongation and the softer O:H always relaxes more than the H−O bond does, $|\Delta d_L| > |\Delta d_H|$. Alternatively, molecules shrink their sizes when they are further separated, and vice versa. Relaxation only changes the segmental length and energy without alternating the nature of the O:H−O or its orientation.

Water absorbs energy through H−O bond contraction and emits energy at its length inversion. The O:H relaxation dissipates energy caped at ~0.2 eV through molecular thermal vibration or even evaporation. The length, energy, and the stretching vibration frequency ($d_x$, $E_x$, $\omega_x$) of the H−O are about (1.0 Å, 4.0 eV, 3200 cm$^{-1}$) and the O:H are about (1.7 Å, 0.2 eV, 200 cm$^{-1}$) at 277 ºK of maximal mass density, $\rho_M$ = 1.0 g/cm$^{-3}$ [51]. The bond angle, segmental length and the O—O repulsion undergo fluctuation.

Fig 1 a inset formulates the relations for XPS and phonon spectroscopies [76, 74]:

i) The O 1s binding-energy shifts proportionally to the H−O bond energy, $\Delta E_{1s} \propto \Delta E_H$, while the $\Delta E_L$ is too small to make significant contribution;

ii) The segmental stretching vibration frequency $\omega_x$ depends functionally on the segmental length $d_x$, energy $E_x$, and its reduced mass $\mu_x$, of the oscillator; and,

iii) The segmental cohesive energy $E_x$ is inversely proportional to its length $d_x$ with $m > 0$ being the bond nature index.

Fig 1b shows the general density dependence of molecular separation $d_{OO}$ and the $d_H \sim d_L$ correlation [77]. The $d_H$ and $d_L$ are projected along the O—O. With the known molecular separation, $d_{OO} = 2.965$ Å, for monolayer water skin instance [54], one can readily derive $d_H = 0.889$ Å, $d_L = 2.076$ Å, and $\rho = 0.75$ g/cm$^3$ compared with $\rho = 0.92$ g/cm$^3$ at the freezing point for bulk water, $T_N = 258$ °K [22].

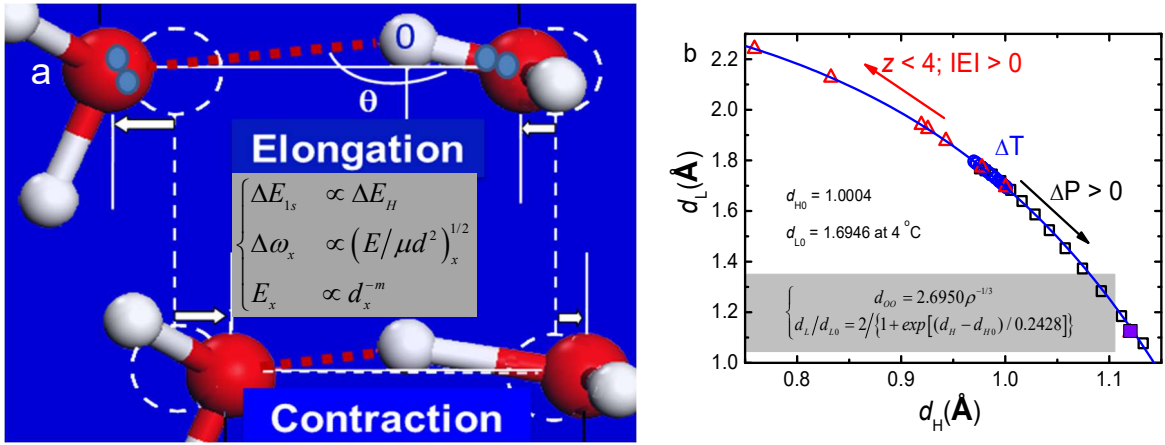

**Fig 1 | O:H−O segmental length cooperative relaxation** [63, 78]. (a) O—O distance changes by shortening one segment and lengthening the other because of coupling interaction. The O:H always relaxes more than the H−O does because of the O:H−O segmental disparity. The primary rule for (b) O:H−O length cooperativity by compression (P), thermal excitation (T), molecular undercoordination (z), and electrostatic polarization (E).

3. Rule 2: Specific-heat disparity versus quasisolidity

− *O:H−O segmental stiffness and energy define its specific heat and multiphase density oscillation* [79]

Because of the segmental disparity in stiffness, introducing the specific-heat of Debye approximation, $\eta_x(T/\Theta_{Dx})$, is necessary for each segmental to describe the response of the O:H−O bonds to thermal excitation [79]. The specific-heat is the energy required to raise the segment temperature by one °K. The thermal integration of the $\eta_x(T/\Theta_{Dx})$ from 0 °K to the evaporation temperature, $T_{Vx}$, at which segment thermal rupture occurs, equals the segmental cohesive energy $E_x$. The phonon frequency $\omega_x$ determines the Debye temperature $\Theta_{Dx}$ through Einstein's relation, $\Delta\Theta_{Dx} \propto \Delta\omega_x$. Therefore, an external perturbation mediates the $\eta_x(T/\Theta_{Dx})$ curves through $\omega_x$ and $E_x$ relaxation. The $\eta_x(T/\Theta_{Dx})$ of lower $\Theta_{Dx}$ approaches to its saturation more quickly. The known $(\omega_x, E_x, \Theta_{Dx}) = \sim(200$ cm$^{-1}$, 0.2 eV, 192 °K) for the O:H and $\sim$(3200 cm$^{-1}$, 4.0 eV, 3200 °K) for the H–O bond in the bulk water derived the $\Theta_{DH}/\Theta_{DL} = 3200/192$, which yields the respective $\eta_x(T/\Theta_{Dx})$ curves, as shown in Fig 2a (broken curves as the standard reference).

The interplay of the segmental $\eta_x(T/\Theta_{Dx})$ determines the thermodynamics of water ice. The superposition results in two intersecting points and reproduces the critical temperatures for boundaries of the known phases displayed under atmospheric pressure [22], see Fig 2b. One may infer from the $\rho(T) - \eta_x(T/\Theta_{Dx})$ correspondence that the segment having a lower $\eta_x$ value follows the regular rule of thermal expansion but the other segment responds to thermal excitation contrastingly, in a "master-slave" manner.

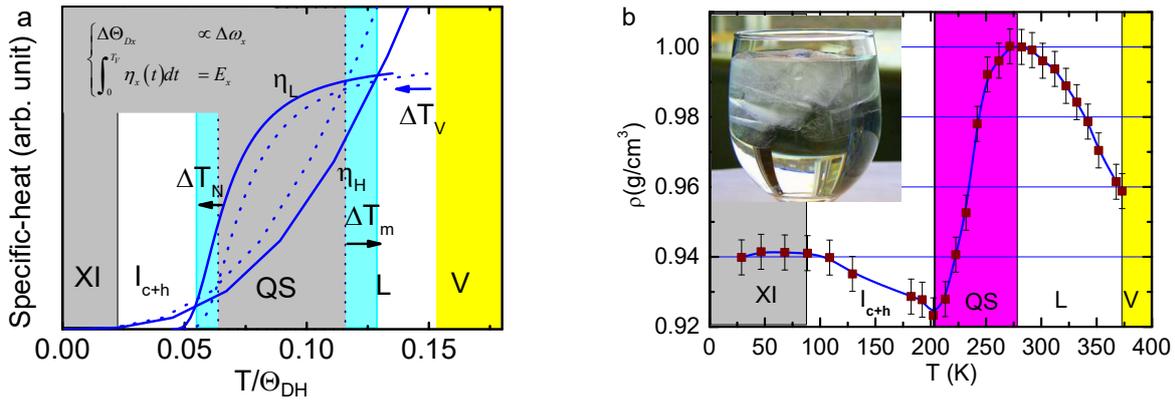

**Fig 2 | Specific-heat disparity and multiphase thermal mass density oscillation** [79]. (a) Intersection points define the quasisolid phase (QS) whose boundaries close to temperatures for homogeneous ice nucleation ($T_N$) and melting ($T_m$). Molecular undercoordination disperses the QS boundary outwardly by H−O contraction and O:H elongation through Einstein's relation. (b) Segmental specific-heat ratio defines the thermal slope of density over all phases for water ice and the critical temperatures vary with volume size at the nanometer scale (T ≥ 273 °K bulk water; T ≤ 273 °K 1.4 nm sized droplet) [21].

Most strikingly, the superposition of the $\eta_x(T/\Theta_{Dx})$ creates the ever unaware quasisolid phase (QS) whose boundaries ($\eta_L/\eta_H \equiv 1$) correspond to extreme densities and close to temperatures for melting $T_m$ (277 °K, 1.0 g/cm$^3$) and freezing $T_N$ (258 °K; 0.92 g/cm$^3$). The $T_N$ shifts from 258 to 205 °K when turning the bulk water into 1.4 nm sized droplet [79]. The $\Delta T_N \propto \Delta T_V \propto \Delta E_L$ and $\Delta T_m \propto \Delta E_H$ relations hold roughly though dispersion of the specific heat curves. In the QS phase ($\eta_L/\eta_H > 1$), cooling shortens the $d_H$ less than the $d_L$ elongates and the volume expands gradually to a maximum at the $T_N$, which clarifies why ice floats. It would be proper to define 277 °K as the $T_m$ because the density profile shows no transition at $T_m$ = 273 °K. Indeed, measurements showed that the (Tm, $T_N$) = (277, 261.2 °K) for the deionized water (54).

In the Vapor phase ($\eta_L \cong 0$), the O:H interaction is negligible, and the gaseous molecule can be taken as an isolated structure unit that has the shortest H−O bond. In the Liquid and $I_{c+h}$ phases ($\eta_L/\eta_H < 1$), $d_L$ cooling contracts more than $d_H$ elongates, so density increases at different rates. Liquid water and $I_{c+h}$ ice follow the regular rule of thermal expansion, but it is in a completely different mechanism. The energy storage by the $d_H$ thermal contraction is prerequisite to the Mpemba effect for energy emission at cooling [55]. In the XI phase, the $\eta_x(T/\Theta_{Dx})$ approaches zero, $\eta_L \cong \eta_H \cong 0$, O:H−O segmental length and energy are insensitive to thermal excitation, $\Delta \omega_x \cong 0$ [80, 81]. The cooling ∠O:H−O angle expansion from 165 to 175 ° lowers slightly the mass density [21].

## 4. Rule 3: BOLS-NEP notion versus supersolidity

*− Undercoordination shortens and stiffens the H−O bond and polarizes lone pairs* [22]

**Fig 3**a inset formulates the bond order-length-strength correlation and nonbonding electron polarization (BOLS-NEP) [82]. Atomic undercoordination mediates the performance of defects, surfaces, and nanostructures by local bond contraction, core and bonding electron entrapment, and nonbonding electron polarization. Bond contraction raises the local charge and energy density and its strength gain deepens the interatomic potential-well that traps electrons accommodated in the core levels and bonding orbitals. In turn, the locally and deeply entrapped electrons polarize those nonbonding electrons of the lower coordinated edge atoms, creating the double layer of polarization at edges or rims of a substance [83]. The BOLS-NEP perturbs the crystal potential of the Hamiltonian, which governs the performance of bonds and electrons in the energetic-spatial-temporal domains, and the structure, morphology, and macroscopic properties of a substance [1]. The BOLS-NEP premise has thus reconciled the unusual performance of undercoordinated adatoms, defects, surfaces, and aqueous and solid nanostructures. The size dependency and derivacy in mechanical strength, thermal stability, electronic and photon emissivity, dielectrics, magnetism, and catalysis has led to the revolutionary in condensed matter chemistry and physics. For instance, the spin-resolved polarization by *sp*-orbital hybridization and atomic undercoordination plays the dominant role in determining the edge and skin superconductivity of topological insulators and monolayer high-$T_C$ superconductors [83].

Likewise, water molecular undercoordination shortens and strengthens the H–O bond but lengthens and softens the O:H nonbond contrastingly and cooperatively of the coupled O:H–O bond associated with strong polarization [84, 4]. H–O bond contraction my not follow exactly the primary law quantitatively because the O—O coupling interaction. Fig 3b shows the O:H−O potential paths as a function of the $(H_2O)_{N\leq 6}$ luster size [85]. Lagrangian resolution to the coupled O:H−O bond oscillation transforms the known $(d_x, \omega_x)$ into the segmental force constant and cohesive energy $(k_x, E_x)$ for each sized $(H_2O)_{N=2-6}$ cluster. Under molecular undercoordination, the O:H−O does relax cooperatively in a "master-slave" fashion as expected, see segmental displacement in Fig 3b and cooperativity in Fig 4d.

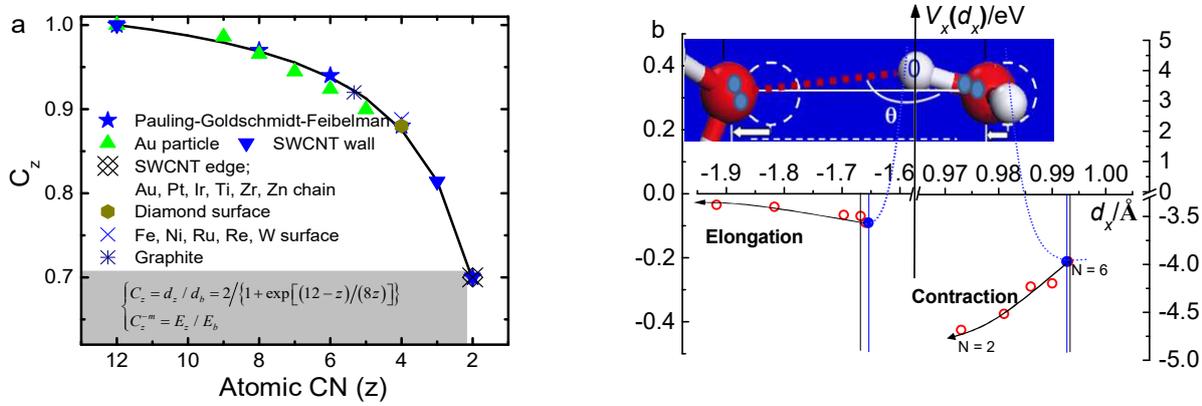

**Fig 3 | Undercoordination resolved bond relaxation and O:H−O potentials** [85, 82, 1]. (a) BOLS formulation of atomic undercoordination-resolved bond contraction ($z < 12$), with $m > 0$ being the bond nature index. (b) O:H–O potential paths for the sized $(H_2O)_{N=2-6}$ clusters ($\Delta d_H < 0$, $|\Delta E_H| > 0$; $\Delta d_L > 0$, $|\Delta E_L| > 0$). The blue dots in (b) are the initial equilibrium for N = 6.

Molecular undercoordination not only disperses the QS boundary outwardly by $\omega_x$ relaxation but also strongly polarizes the QS phase, leading to the skin supersolidity that is hydrophobic, less dense, lubricate, mechanically and thermally more stable, diffusive, and viscoelastic. The supersolidity is extended from the elastic and repulsive contacting interface between Helium fragments that undergo frictionless motion at mK temperatures [86]. Existing throughout the volume, the QS arises from specific-heat disparity but the supersolidity results from polarization by molecular undercoordination. The supersolidity enhances the QS phase at bonding network ends such as defects, skins, droplets, hollow bubbles and skins of bulk species, regardless of the structure phase and thermal insensitive.

Quasisolid undergoes cooling volume contraction and the supersolidity is subject to undercoordination expansion. Droplet size reduction increases the fraction of undercoordinated molecules and reduces the effective molecular *CN* of the skin. The droplet-size-induced $\Theta_{Dx}(\omega_x)$ relaxation mediates the specific-heat and hence disperses the extreme-density temperatures or boundaries of the QS. The QS dispersion by undercoordination lowers the $T_N$ and $T_V$ and raises the $T_m$, leading to the "superheating" and "supercooling" phenomena observed from bubbles and droplets.

5. Multifield perturbation: force-field O:H−O length cooperativity

*− Perturbation relaxes the O:H−O segmental length cooperatively in a 'master-slave' manner* [63]

The O:H−O cooperative relaxability is proven universally true. Fig 4 shows results computed using the COMPASS'1998 force-field of H Sun [87] using a complex unit cell of 64 molecules for (a-c) under perturbation. Indeed, the relaxation proceeds in a 'master-slave' manner. The arrows denote the master segments and their shifting directions under perturbation of the given degrees of freedom. A stimulus dislocates both $O^{2-}$ anions in the same direction but by different amounts. The softer O:H (upper part) always relaxes more than the stiffer H−O with respect to the $H^+$ coordination origin.

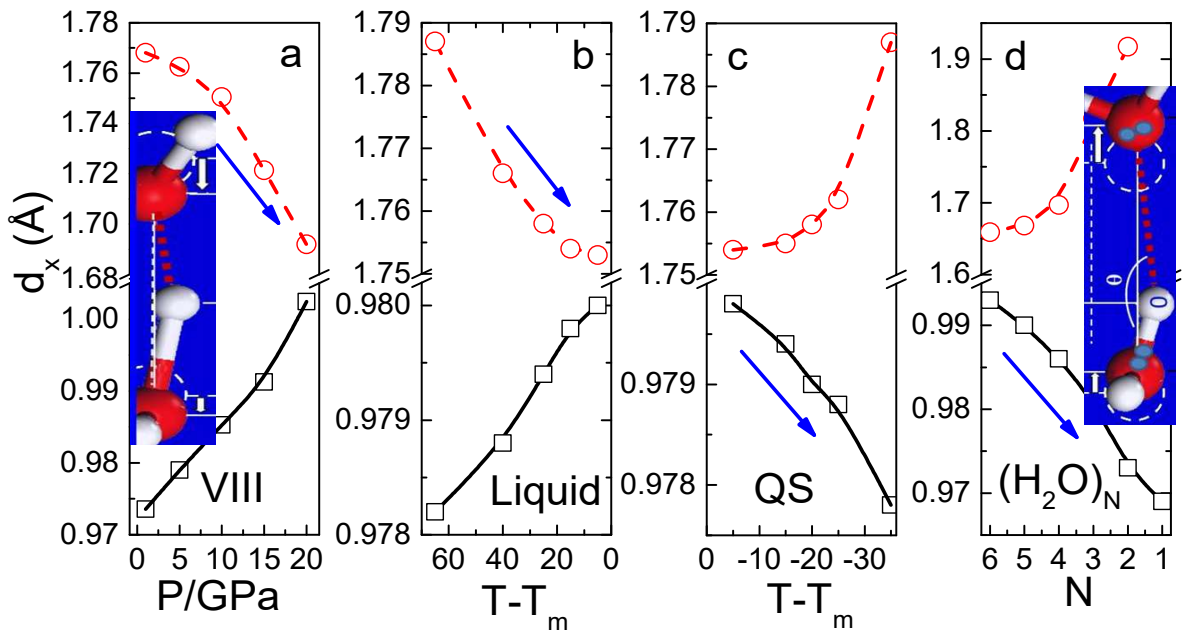

**Fig 4 | O:H–O bond segmental length cooperative relaxation.** O—O repulsion dislocates O ions in the same direction by different amounts (insets) under (a) mechanical compression, cooling of (b) Liquid and (c) Quasisolid (QS) phase (unit in °C), and (d) undercoordination by reducing the $(H_2O)_N$

size from N = 6 to 2. Arrows denote the master pieces and their relaxation directions. The H–O bond always relaxes less than the O:H and both of them relax contrastingly in their curvatures and slopes, irrespective of the applied stimulus or the structural order because of the O–O Coulomb repulsive coupling (reprinted with permission from [63]).

Panels a shows the effect of compression, panels b and c are effect of cooling of liquid and QS phase and panel d the sized $(H_2O)_{2-6}$ clusters [63]. The slopes and the curvatures of the coupled relaxation curves in each panel follow the $(dd_L/dq)/(dd_H/dq) < 0$ and $(d^2d_L/dq^2)/(d^2d_H/dq^2) < 0$ with $q$ being the parameter of perturbation. The ∠O:H−O angle relaxation only contributes to the crystal geometry and mass density. O:H−O bond bending has its own mode isolating from the H−O or the O:H stretching vibrations [63], which is the advantage of a spectroscopy.

Results confirmed that (a) mechanical compression and (d) molecular undercoordination effect oppositely on the O:H−O relaxation and that the O:H−O responds to cooling contrastingly in (b) the liquid and (c) the QS phase, confirming the O:H−O cooperativity theoretical predictions. Raman spectroscopy investigations [79, 75, 22] confirmed the corresponding O:H−O bond stiffness relaxation. Preliminary results showed the efficiency and essentiality of the BOLS-NEP and coupled O:H−O bond theories in dealing with water ice, which opened the entrance and paved the path directing to the core physics and chemistry of water and ice.

6.  DFT derivatives: site and orientation resolved O:H−O relaxation

*− O:H−O segmental length and stiffness are rather sensitive to the coordination environment* [53]

Fig 5 shows the DFT calculated O:H−O length and vibration frequency (stiffness) cooperativity in ice skin [53]. The spectra are derived from a unit cell of 64 molecules with and without a vacuum slab, as shown in Fig 5 a inset. The spectral difference shows the abundance (integral), extent (peak shift), and fluctuation (peak width) transition of the segmental length and stiffness from the bulk (B) to the skin (B) and the dangling H−O radical (R). As expected, the $d_H$ contracts from the bulk value of ~1.00 to ~0.95 and 0.93 Å when move from the bulk to the skin and radicals. The $d_L$ elongates from the bulk value of ~1.68 to the skin of ~1.90 Å with high fluctuation. The corresponding □$_L$ and □$_H$ transit from the bulk to the skin and the free H−O radicals contrastingly. the P peak arises from the screening and splitting of the crystal potentials by the polarization in numerical derivatives.

Wang and co-workers [88, 45] examined using DFT calculations the site and orientation resolved electronic binding energy and H−O stretching vibration for the sized $(H_2O)_n$ clusters (n = 17, 19, 20, 21, 23 and 25). The H−O bonds are classified into five groups according to their coordination environments: the dangling H−O bonds (D), the H−O of the O:H−O formed between the dangling $H_2O$ molecules (C), those of the O:H−O formed between the $H_2O$ molecules without dangling H−O bonds, and those of the O:H−O bond between the tetrahedral-coordinated $H_2O$ and its neighboring molecules. The calculated spectra in Fig 6 revealed that the dangling bond D peak keeps constant at 3760 cm$^{-1}$ throughout the considered sizes, while the frequencies of other peaks are cluster size dependence. Observations confirmed the effect of coordination environment on the H−O bond length contraction, energy gain, and its vibration frequency blueshift when moving from the f core center to the rim dangling bond of the clusters [63].

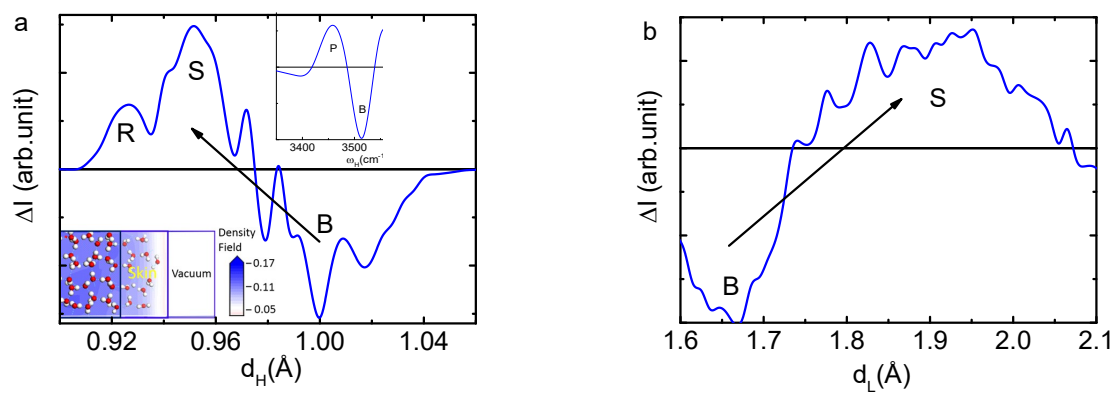

**Fig 5 │ Skin O:H–O segmental length and stiffness cooperative relaxation** [53]. Length and stiffness (inset) transition for the (a) H–O bond and (b) O:H nonbond from the bulk value (B) to the skin (S) and to the H–O free radicals (R). Inset (a) shows the complex unit cell denoted with bulk, skin, and a vacuum slab. The P components arise from the screening and splitting of the crystal potentials by polarization.

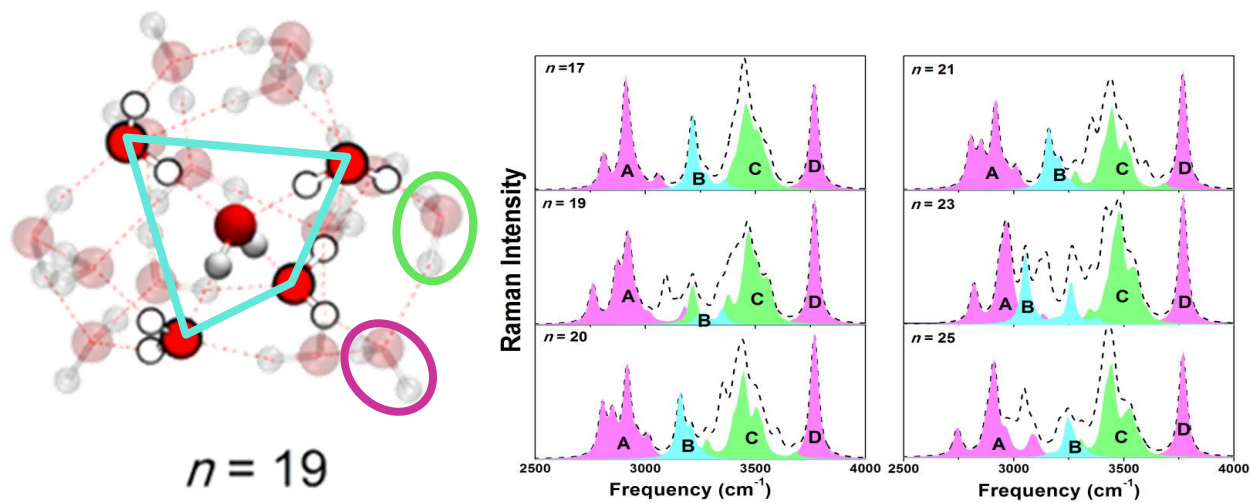

**Fig 6 | Computational H−O stretching vibration modes in the (H₂O)ₙ clusters** [88]. The black dashed lines convolute the H−O vibration modes of the entire clusters. The sharp feature D corresponds to the H−O dangling bonds, C to the H−O of the O:H−O bonds between molecules at rims. Features A and B to the H−O bonds inside the clusters. Reprinted with copyright permission from [88].

7. Skin phonons: H−O bond stiffening and O:H softening

*− DPS resolves O:H−O phonon transition from the mode of ordinary water to its supersolid* [74]

**Fig 7** compares the differential phonon spectroscopy (DPS) $\omega_H$ for skins of 298 °K water and (153-258) °K ice [65] and the $\omega_{DO}$ for the (a) nanosized (H+D)₂O droplets [33]. Mixing the D₂O and H₂O in the spectroscopy aims to avoiding impurity signatures appeared in the H−O phonon band. The characteristic D−O vibration frequency differs from that of the H−O frequency because the isotope effect on the reduced oscillator masses. The DPS is the difference between two spectra collected at different polar angles from the surface or from sized water droplets, upon all the spectral peak areas being normalized [89].

Results confirmed again that molecular undercoordination does shorten the H−O and the D−O bond and stiffen their phonons. Strikingly, skins of water and ice share an identical $\omega_H$ of 3450 cm⁻¹. The DPS peak integrals suggest that the skin of ice is 9/4 times thick of the liquid, because of thermal fluctuation. The identical $\omega_H$ says that neither a layer of ice (3150 cm⁻¹) covers the liquid nor a liquid overlayer (3200 cm⁻¹) stays on ice, instead, skins of both liquid and ice share the same supersolidity of identical H−O bond of 3450 cm⁻¹ vibrating frequency.

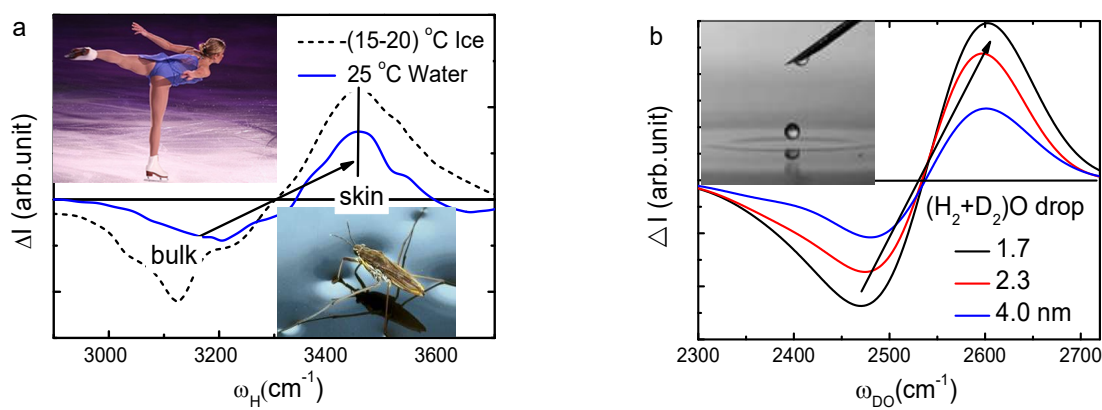

**Fig 7 | H−O stiffness transition from the bulk to the skins of large volumes and nanodroplets** [74, 53]. (a) Skins of 298 °K water and (253-258) °K ice [65] share an identical $\omega_H$ of 3450 cm⁻¹ and

the (b) D–O phonon [33] transits from below 2550 to its above for the skin of droplet. Insets show ice slipperiness, water skin toughness and the elasticity and hydrophobicity of skins of droplet and water.

One can estimate the skin shell thickness of the core-shell structured droplet. Integrating the DPS peaks in **Fig 7** b gives rise to the frequency-, or size-dependent fraction of D–O bonds transiting from the mode of water in the core to the skin, $f(D) = 0.54D^{-1}$. The fraction coefficient equals the skin-to-volume ratio of a spherical droplet of $V \propto R^3$, which yields, $f(D) = \Delta V/V = 3\Delta R/R$. The skin thickness $\Delta R_{skin} = f(D)D/6 = 0.90$ Å equals the H–O dangling bond length [63] featured at 3610 cm$^{-1}$ [22]. The DPS distills the most outstanding O:H–O monolayer thickness of 2.965 Å [54] and the relaxation gradually converges to the core of the droplet.

SFG measurements [35] uncovered the site and orientation resolved H−O bond vibrating frequencies on the outmost two molecular layers of ice $I_h$ (0001) surface. The frequency of the H−O bond (H−$O_{B1}$) pointing from the first to the second sublayer is above 3270 cm$^{-1}$ and the H−$O_{B2}$ from the second to the first layer is below 3270 cm$^{-1}$ because of the coordination environment. This discovery is consistent with the DFT derivatives [90] and the present BOLS expectation [91, 85]. The less coordinated H−$O_{B1}$ is shorter and stiffener than the H−$O_{B2}$. The frequency of the dangling H−O bond is at 3700 cm$^{-1}$ measured by SFG and 3610 cm$^{-1}$ by Raman reflection under the ambient temperature.

The skin supersolidity is responsible, as Fig 7 insets (a) illustrated, for the skin toughness of water, slipperiness of ice, and the mechanical strength, thermal stability, elasticity, hydrophobicity and fluidity of water droplet traveling in a microchannel. The supersolidity of the skins of the droplet and bulk water endows the droplet bouncing rounds when it falls on water (inset b).

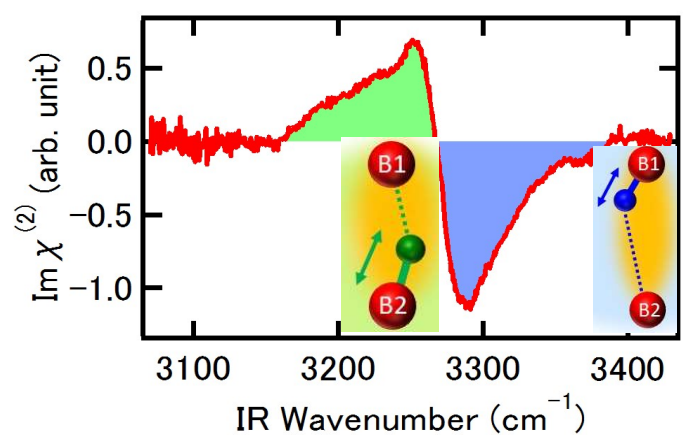

**Fig 8 | Site and orientation resolved SFG H–O vibration frequency.** Undercoordination resolves the frequencies of the $O_{B1}$:H−$O_{B2}$ and the $O_{B1}$−H:$O_{B2}$ bonds (inset) between outermost two sublayers of the ice $I_h$(0001) skin [35]. Insets illustrate segmental bond lengths, orientations, and frequencies of the H–O stretching vibrations. The positive peak (< 3270 cm$^{-1}$) corresponds to the H−$O_{B2}$ vibration

(shaded in green) and the valley (> 3270 cm$^{-1}$) to the $O_{B1}$−H (shaded in blue). The less coordinated $O_{B1}$−H is shorter and stiffer and its H:$O_{B2}$ is longer and softer than the $O_{B1}$:H−$O_{B2}$.

8.  Skin electrons: entrapment and polarization

*− Skin H−O contraction entraps the core and bonding electrons and polarizes lone pairs* [76]

The energy level of an isolated atom shifts deeper when a large volume is formed as the interatomic interaction comes into play. The enenrgy shift is proportional to the single bond energy, according to tight-binding approximation with ommiting the tiny overlapping integral betweeen the same core orbits of adjascient atoms [92]. Atomic undercoordnation shifts further the core level because of the spontenous bond contraction. XPS and UPS data in Fig 9 confirmed that molecular undercoordination induces local O 1 $s$ binding energy entrapment and nonbonding electron polarization. The O $1s$ level shifts from 536.6 eV to 538.1 and to 539.7 eV when one moves from the bulk to its skin and to monomers in its gaseous phase [30, 31]. The O $1s$ shift fingerprints directly the H−O energy and length change becase of the $E_H \propto d_H^{-m}$ relation.

Near edge X-ray absoprtion fine structure (NEXAFS) revelead the 535 eV predge and a 531 eV gaseous $O_2$ peak for high-density oxygen infilled nanobubbles [93]. The peaks result from the enenergy diffreence between the inner O $1s$ level and the uppermost level [89], $\Delta E_{XAS} = E_v – E_{1s}$ = 531 and 535 eV for $O_2$ gas within the bubble and the liquid $H_2O$, respectively. The $\Delta E_{O2} – \Delta E_{H2O}$ = - 4 eV indicates that the uppermost $E_V$ level and the $E_{1s}$ for $O_2$ gas are deeper than correponding ones of liquid $H_2O$.

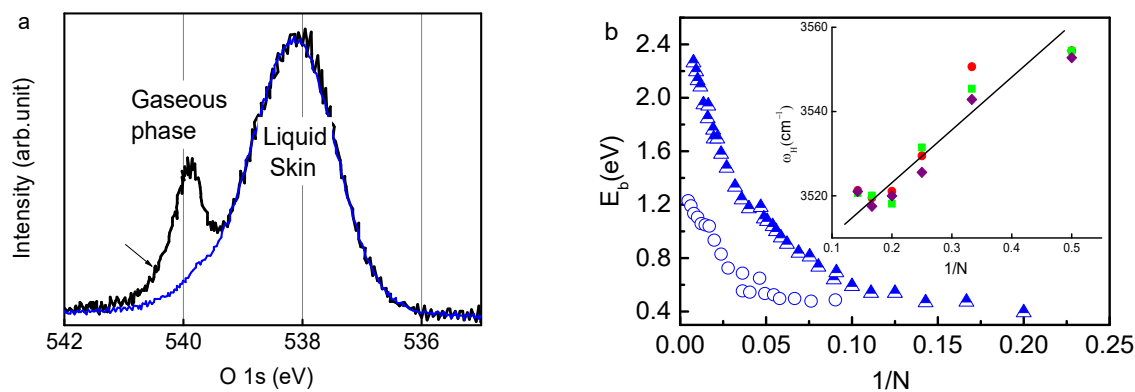

**Fig 9│Undercoordination induced quantum entrapment and polarization** [67, 63, 68, 64, 31]. (a) The O 1s energy shifts from the bulk value of 536 to 538 for the skin and to 540 eV for gaseous state. (b) The hydrated nonbonding electron shifts its energy from the core value of 2.4 and the skin of 1.2 to the limit of 0.4 eV for N = 5 cluster. Inset b shows the cluster size dependent $\omega_H$ blueshift [4].

Due to the local polarization, an oxygen atom gains net charge from -0.616 to -0.652 e when moves from the bulk to the skin [53]. This charge gain enhances further the O—O repulsion at the surface. A free electron injected into water serves as a probe to the local environment without changing the solvent geometry. The hydrated electron will be trapped by the locally oriented $H_2O$ molecules, forming a (-)·$4H_2O$ motif in contrasting polarity of the $Na^+·4H_2O$ [94]. The ultrafast pump-probe liquid-jet UPS [67, 68, 32, 64] probed, see Fig 9b, that the bound energies (equivalent to work function) of the hydrated electrons are centered at 2.4 eV in the bulk interior and at 1.2 eV when they are located at the skin. The bound energy decreases further with the $(H_2O)_N$ cluster size toward a limit of 0.4 eV for $N =$ 5. Inset b shows the $(H_2O)_N$ size dependence of the H−O phonon frequency, agreeing with electronic energy entrapment – smaller droplet size has high fraction of shorter and stiffer H−O bonds which deepens the O $1s$ energy level and enhances the polarization.

## 9.  Ultrafast spectroscopy: electron and phonon lifetime

*− Viscoelasticity and dipolar repulsivity retard carrier dynamics mainly by reflection* [89]

Electron or phonon lifetime $\tau$ probed using ultrafast spectroscopy fingerprints the energy or abundance dissipation dynamics determined by the specific coordination environment. Generally, the phonon lifetime is proportional to its frequency [69] and the electron lifetime is inversely proportional to its bounding energy [32]. The skin high-frequency H−O or D−O phonons and the skin low-bound-energy electrons decay slowly than they do in the bulk.

Fig **10** a shows the D−O phonon lifetime $\tau$ for water droplets confined in the reverse micelles compared with that of the ionic hydration volume for the concentrated NaBr solutions [33]. The $\tau$ increases from 2.6 ps for the bulk water to 3.9 and 6.7 ps as the water transits into the solution with concentration increasing from 32 to 8 $H_2O$ per NaBr solute. The lengthening of the phonon lifetime is proportional to the number fraction of bonds being polarized in the ionic hydration volume. In contrast, the phonon lifetime increases from 2.6 for the bulk to 18 and 50 ps for droplet of 4.0 to 1.7 nm size [33]. The lifetime is decomposed as the effects of polarization and the skin reflection [95].

Fig **10** b shows the droplet size and site resolved lifetime of a hydrated free electron. The internally solvated electron bound energy in a $(D_2O)_{50}^-$ cluster is centered at –1.75 eV and the surface localized electron has bound –0.90 eV. These two states vary with the cluster size and change from $(D_2O)_{50}^-$ to $(H_2O)_{50}^-$. The hydrated electrons live 100 ps longer and almost constant near the surface than those inside the bulk.

One may extend the processes of photofluoresce [96] and very-low-energy electron diffraction by the complex surface potential barrier (SPB) to the ultrafast processes of electrons and phonons [89]. The SPB has two parts, $V(r, E) = ReV(r) + iImV(r, E)$. The real elastic $Re(V)$ part and imaginary inelastic $ImV(r, E)$ part are correlated by Poisson's equation, $\nabla^2 Re(V) \propto \rho(r) \propto ImV(r, E)$, and $-\nabla Re(V) \propto \varepsilon$ (electric field). Interaction with the integral of the inelastic $ImV(r, E)$ reduces the amplitude A of electron beam and interaction with the integral of the elastic $ReV(r)$ shits the phase $\varphi$ of the electron waves traveling in the charge occupied region and its vicinity, $\psi(r, t) \approx Ae^{i(kr-\omega t+\varphi)}$. The $ImV(r, E)$ absorbs energy by activating secondary electrons. Likewise, the complex dielectric constant, $\varepsilon_r(\omega) = \varepsilon_r(\infty) + i\varepsilon'_r(\omega)$, dictates the lifetime of fluorescent light. The $\varepsilon'_r(\omega)$ absorbs photon energy by electron polarization to produce secondary photons.

The viscosity forms the imaginary part and the elasticity is the real part of the complex rheology for sliding friction [6] and phonon wave propagation. The local charge density of the pinned dipoles forms the $ImV(r, E)$ and its image potential produces the $ReV(r)$ for the hydronated electrons waves. In the skin, the high viscosity and charge density depress the respective wave amplitude by absorbing energy, which should shorten the lifetime of the traveling waves, instead. The longer lifetimes of electrons and phonons suggest that the high mechanical and electrostatic elasticity of the supersolid skin reflects waves to form quasi-standing waves with prolonged lifetimes. The higher the elasticity, the stronger reflectivity of the waves. Waves have similar amplitudes and wavelengths and moving in opposite directions could form a standing wave. The lifetime of an ideal standing wave is infinity. The longer electron lifetime of the $(D_2O)_n$ clusters suggests their higher skin reflectivity or lower charge density. The isotope effect adds electron-phonon interaction as another degree of freedom.

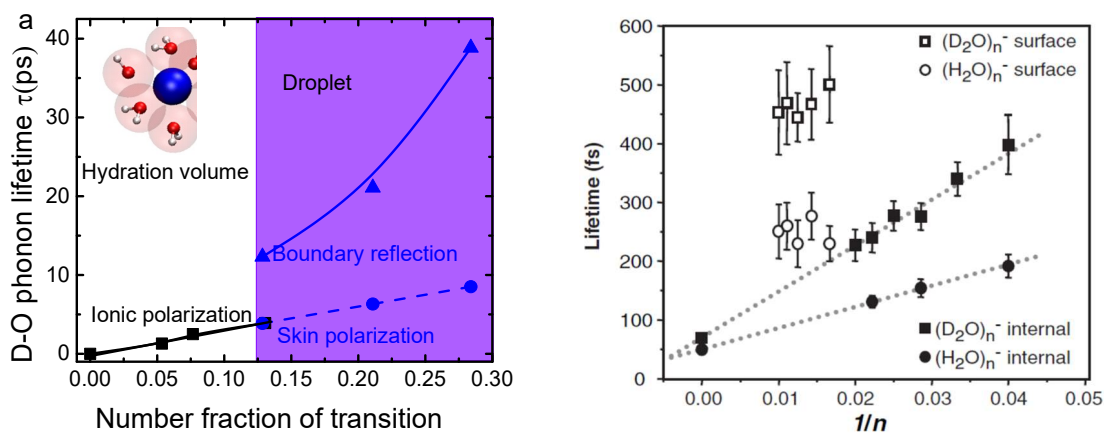

**Fig 10 | Ultrafast phonon and hydrated electron processes of water droplets** [33, 32]. Lifetime of (a) the D−O phonons in the droplet skin compared with that in the $NaBr/H_2O$ solutions [33] and the lifetime of (b) electron hydrated by the sized $(H_2O)_n$ and $(D_2O)_n$ clusters and their skins [32]. Insets (a) illustrate the ionic hydration and the core-shelled water droplet. The skin supersolidity hinders the motion dynamics of both phonons and nonbonding electrons [55].

## 10. Ice slipperiness and water skin toughness: elasticity and repulsivity

*− Low-frequency O:H phonon elastic adaptivity and dipolar repulsivity entitle skin mechanics* [48]

Ice slipperiness means non-sticky and frictionless motion of a body sliding on ice, see Fig 7a inset. The friction coefficient is μ = 0.005 ~ 0.1 for a steel-pin on an ice-disc but the μ of ice on ice varies from 0.05 (at 253 °K) to 0.5 (271 °K) [97]. Besides pressure melting [23] and frictional heating [98] that have been ruled out, mechanisms of molecular undercoordination [99], supercooling [36], low-frequency molecular vibration [97], molecular rolling [36], skin nanorheology [6], and the presently described skin supersolidity [48] have advanced the understanding of ice friction from various perspectives. Spectroscopy measurements using LEED, NMR, SFG, and proton backscattering suggested the following:

i) surface molecules rotate five orders in frequency at 273-293 °K than they do in the bulk [34].
ii) surface oxygen vibrates up-down more rapidly than in the plane [27].
iii) surface oxygen vibrates 3.3 times in amplitude of its bulk value [28].
iv) surface molecules roll readily to serve as bearing [36].

Mechanical detection suggests that the rheological skin of $10^2$ nm thick has an elastic and a viscous component, which dictates the scratching friction [6]. At the atomic scale, Krim [97] proposed that interface lattice vibration and charge distribution play significant roles in sliding friction. Atomic vibration at a surface creates phonons with certain distinct frequencies. If the "plucking" action of atoms in the opposite surfaces of contacting motion, phonon resonance of both surfaces raises the friction coefficient, which explains why ice on ice has a higher friction coefficient.

Contrastingly, it was thought that an ice-like layer makes water skin tough, elastic and hydrophobic, as confirmed using SFG spectroscopy and MD calculations [100]. A falling water droplet bounces rounds on water surface before it disappears, see Fig 7b inset, which straightforwardly shows the high elasticity and hydrophobicity of both skins regardless of their curvatures.

Computations [55] suggested that the skin stress and viscosity increase with the number reduction of its molecular layers. The stress increases from 31.5 to 73.6 mN/m when transiting a film from 15 to five-layer of molecules, which approaches to the measured value of 72 mN/m for water skin at 298 °K. The skin viscosity increases from 0.007 to 0.019 $\times 10^{-2}$ mN·s/m$^2$, agreeing with the trend of measurements [6]. Theoretical reproduction of the stress thermal decay of water skin, derived the $\Theta_{DL}$ = 192 °K and the $E_L$ = 0.095 eV [101, 47].

Does a liquid-like skin cover ice and an ice-like layer form on liquid? Results in Fig 5 - Fig 11

demonstrate the skin supersolidity of nanodroplets, bulk water and ice, characterized by the identical 3450 cm$^{-1}$ H−O phonon frequency and nonbonding electron polarization. The frequency transition from 200 to 75 cm$^{-1}$ and the high amplitude of the softer O:H phonon entitled water molecules with high elastic adaptivity to sliding friction. The polarization offered the skin with electrostatic repulsivity.

The counterpart of ice friction is always negatively charged regardless of the nature of its material because the undercoordination induced local polarization by the densely entrapped core electrons [1]. Such an elastic adaptive and repulsive contacting interface not only lowers the effective contacting force but also prevents charge from being transport between the counterparts. For ice on ice, O:H phonon resonance and interface fusing regelation raise the friction coefficient [97]. Therefore, the softer O:H phonon endows the elastic adaptivity and the nonbonding electron polarization ensures the hydrophobicity and repulsivity, making ice slippery and water skin tough. The skins and water and ice share the common gel-like supersolidity of elastic, viscus, less dense, repulsive, thermally stable.

## 11. Thermal stability: H−O bond stiffening and O:H softening

*− Heating can hardly shorten the undercoordination-shortened H−O bond further* [57]

The full-frequency Raman spectra in Fig 11a confirmed the thermal H−O stiffening and O:H softening of liquid water because of the lower the specific heat $\eta_L$ than the $\eta_H$ [57]. O:H thermal expansion shortens the H−O by O—O repulsion. The convolution of the bulk component centered at 3239, skin at 3443, and H−O radical at 3604 cm$^{-1}$ forms the $\omega_H$ band for 278 °K water. The H−O radical $\omega_{H-O}$ has a maximum at 300 °K and then undergoes thermal softening by -0.40% at 370 °K because of the absence of O—O coupling that ensures H−O thermal contraction. Heating from 278 to 340 °K stiffens the $\omega_{skin}$ by 0.38% and the $\omega_{bulk}$ by 1.48%. At T > 340 °K, the skin H−O bond turns to be thermal elongation because of weakening of the O—O repulsion by thermal depolarization. The skin H−O bond is thus proven thermally more stable than it is in the bulk. The shortened H−O bonds can hardly be further shortened by thermal perturbation.

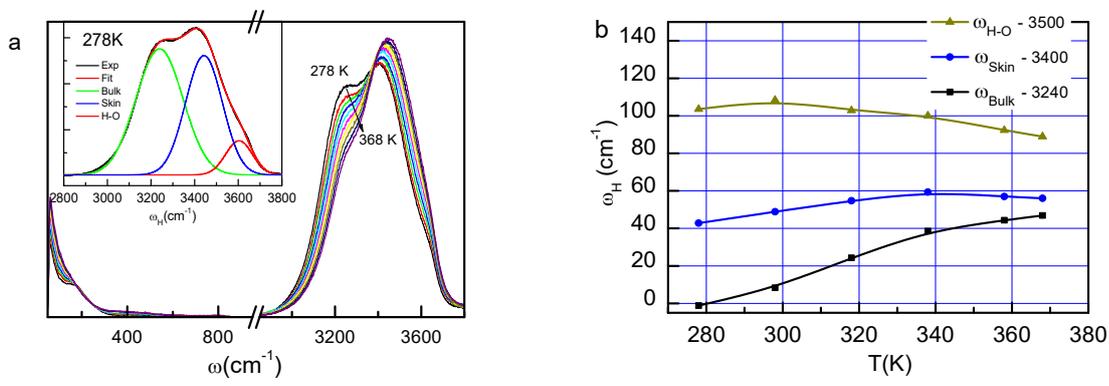

**Fig 11 | Supersolid skin thermal stability** [57]. Temperature dependence of (a) the full-frequency Raman spectra and (b) the frequency shift of the bulk, skin, and H–O dangling bond components. The H–O dangling bond undergoes thermal expansion (redshift) at T > 300 °K. Both the bulk and the skin components undergo thermal contraction, at different slopes. At T > 340 °K, the skin component turns to be thermal elongation, showing the O—O repulsive weakening.

12. Thermal transport: skin specific-heat and thermal-diffusivity

*− O:H−O bond memory and skin supersolidity resolve the Mpemba's heat transport dynamics* [55]

One typical phenomenon in the fluid thermal transportation is the Mpemba effect, firstly noted by Aristotle in the 350 B.C. and asserted by, and named after, Mpemba in the 1960's [56]. Under the same cooling condition, hot water cools faster than its cold though the successful observation is infrequent. This phenomenon has been attributed to convection, evaporation, impurity, supercooling, geometric structure, etc.

The Mpemba effect integrates the energy "heating absorption, cooling emission, cross-skin transportation, and source-drain interface dissipation" dynamics of liquid water [55]. One must consider these processes as a collection when dealing with this problem. Incorporating the skin supersolidity into the boundary and initial condition problem of Fourier fluid thermodynamics has reproduced the signatures of observations, see Fig 12 [55]. One characteristic is the interception of the $\theta(\theta_i, t)$ cooling curves from different initial-temperatures $\theta_i$. The other is the temperature difference between the skin and the volume of the liquid water, $\Delta\theta(\theta_i, t)$ [102]. The liquid temperature $\theta$ decays exponentially ($t$) with a characteristic lifetime $\tau_i$. The curve of higher initial temperature has a shorter $\tau_i$, and vice versa. The skin becomes warmer than its volume before reaching their equilibrium. The $\Delta\theta(\theta_i, t)$ is proportional to the $\theta_i$. The simulative expression in Fig 12a gives the $\theta_i$-resolved rate of the temperature decay, $d\theta(\theta_i, t)/dt = \theta(\theta_i, t)/\tau_i$, showing that the warmer liquid cools more quickly.

Quantitative reproduction of the observed characteristics revealed four facts pertained to this phenomenon:

1) **O:H−O bond memory**. Only could the coupled O:H−O bond in liquid phase absorb energy by H−O heating contraction and emits energy by its inverse. The rate of energy emission is proportional to its initial storage, or the extent of H−O thermal contraction from its equilibrium, $\Delta E_H \propto (d_H - d_{H0})^2$. The rate of energy emission, $dE_H/dt \propto (d_H - d_{H0}) \times dd_H(\theta_i, t)/dt$, and, $dd_H(\theta_i, t)/dt = [dd_H(\theta_i, t)/d\theta(\theta_i, t)] \times [d\theta(\theta_i, t)/dt]$. One can obtain the $(dd_H(\theta_i, t)/d\theta)$ from the $\rho(\theta)$ profile and the $d\theta(\theta_i, t)/dt$ from equations in Fig 12 a. This fact shows the O:H−O bond memory that is absent from a system without the lone pairs or the strong coupling between intermolecular and intramolecular interactions.

2) **Skin high thermal diffusivity**. Fourier equation has a term of thermal diffusion with the coefficient of $\alpha = \kappa/(\rho C_p)$ and a term of convection at $10^{-4}$ m/s velocity. The $\alpha$ is proportional to the thermal conductivity $\kappa$ and inversely proportional to the mass density $\rho$ and the constant-pressure specific-heat $C_p$. The cross point in Fig 12a can only be reproduced by taking the skin mass density $\rho = 0.75$ g/cm$^3$ in calculation regardless of the term of convection. The $\kappa/C_p$ was considered to change insignificantly from the core to the skin. The supersolid skin has a high thermal diffusivity because of its lower mass density.

3) **Skin low specific heat**. Because the thermal flux Q crossing the skin-core interface conserves, $Q = (\theta C_p)_{skin} = (\theta C_p)_{core} \geq 0$, or, $\theta_{skin}/\theta_{core} = C_{p,core}/C_{p,skin}$. The temperature difference, $\Delta\theta(\theta_i, t) = \theta_{skin} - \theta_{core} = \theta_{core}(C_{p,core}/C_{p,skin} - 1) \geq 0$, which yields, $C_{p,skin} < C_{p,core}$, as it can be seen from Fig 2a the specific-heat curve predictions - the liquid specific-heat for the undercoordinated H−O is lower than its fully- coordinated reference.

4) **Non-adiabatic heat dissipation**. Improper source-drain interface conditions such as thermal insulation or a too large source volume will prevent the phenomenon from being take place. Other extrinsic factors such as evaporation, impurity, and supercooling play negligible roles for the Mpemba effect, as only thermal diffusivity and convention are involved in the Fourier fluid thermodynamics.

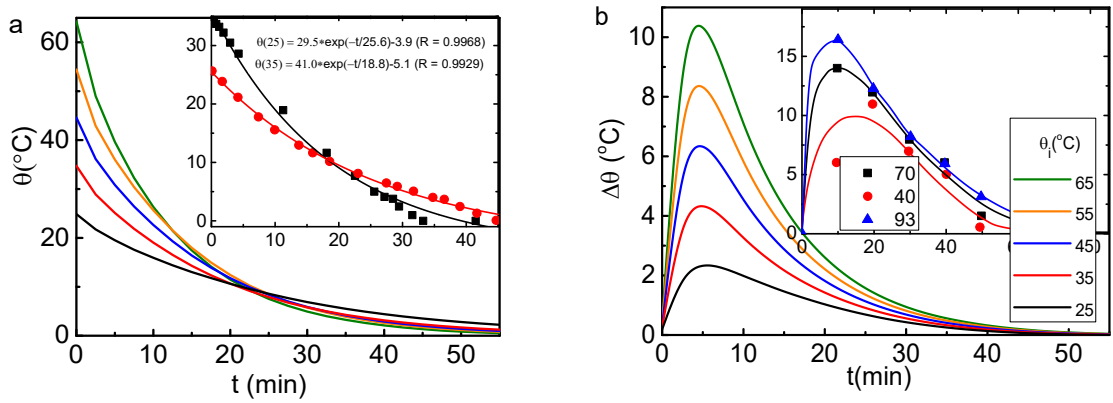

**Fig 12 │ Skin thermal-diffusivity elevation and specific-heat depression** [102, 55]. Numerical reproduction of the measured (insets) initial-temperature dependence of (a) the $\theta(\theta_i, t)$ decay and (b) the skin-bulk temperature difference $\Delta\theta(\theta_i, t)$ of warm water. High skin thermal diffusivity due to density loss ensures the characteristic intersection. (b) The $\Delta\theta(\theta_i, t)$ arises from the skin lower specific heat because heat flux conserves at the interface.

Mpemba effect can only be reproduced with involvement of the supersolid skin regardless of the presence of convection. O:H−O cooperative relaxation lowers the skin mass density and its specific-heat, which raises the thermal diffusivity to favor heat outflow from the liquid; the H−O thermal contraction absorbs energy; O:H−O memorability and recoverability ensures the unusual manner of heat emission. The Mpemba effect, or its inverse, should happen only to systems with involvement of strong coupling of the intermolecular bonding and intramolecular nonbonding lone pair interactions, which ensures the low mass density and high skin supersolidity, bond memory, and heating bond energy absorption. Contrastingly, most known materials have high skin mass density because of the atomic undercoordination induced bond contraction and the bond is subject to energy emission by thermal expansion, instead [1].

13. Ultrafine bubble and droplet: durability and reactivity

− *Supersolidity raises the viscoelasticity and the reactivity by polarization* [76]

Nanobubbles (< 200 nm in diameter) can form easily by dissolving gases like argon, hydrogen, nitrogen, oxygen, and methane in bulk water [10]. According to the classical view of the air–water interface, such bubbles should not exist at all. The small radius of curvature implies a high Laplace pressure inside the bubble that should drive gas diffusion across the interface and cause the bubbles to dissolve almost instantly [103]. However, nanobubbles have peculiar properties such as long lifetime and high gas solubility owing to their negatively charged surface and high internal pressure [93]. Being stable for days or even for months than classical theory can predict, nanobubbles are used in fields such as diagnostic aids, drug delivery, water treatment, biomedical engineering, degradation of toxic compounds, water disinfection, and cleaning of solid surfaces including membrane [104].

The unexpected stability was thought an awkward but conspicuous instance of "surface misbehavior". The liquid/gas interface resists mass diffusion. Theoretical investigation [60] suggests that the limited gas diffusion and the pinned contact line of the nanobubbles lead to the slow dissolution rate.

A bubble is just the inversion of a droplet. A hollow sphere like a soap bubble contains the inner and the outer skins of different curvatures [105]. Both skins are in the supersolid state and the volume fraction of such supersolid phase over the entire liquid-shell volume is much greater than simply a droplet. Nanobubble and nanodroplet have high skin curvatures, $\pm 1/K$. The effective molecular $CN$ varies with the skin curvature, $z_{cluster} < z_{droplet} < z_{flat} < z_{cavity} < z_{bulk} = 4$. The geometrical configuration of the skin molecules stays the bulk attribute, but the length and energy vary with the $CN$ loss - smaller molecular size but larger separation. The fraction of supersolid molecules increases with the curvature. For a sufficiently small water droplet or a bubble, the volume of the skin and the volume of the core is compatible, they hold a bi-phase structure of low-density skin and high-density core of the same tetrahedral structure. If the structure is sufficiently small the skin supersolidity becomes dominant. So, the extent of supersolidity becomes more pronounced as the droplet size is reduced.

Therefore, bubbles show more significantly the supersolidity nature – elastic, hydrophobic, and less dense, which makes bubbles mechanically stronger, chemically more active, and thermally more stable. The strong skin polarization prevents gas diffusion across the skin of the bubble. The supersolidity is a notion that appeals to old ideas about hydrophobic particles creating a highly ordered and ice-like hydration shell in aqueous solution.

The intrinsic behavior of the O:H–O between the undercoordinated molecules is the key controlling the performance of small bubbles. The skin supersolidity stems the unusual thermal and mechanical stability of the curved surfaces. Therefore, nanoscale water ice is more elastic, hydrophobic, repulsive, and less dense with even lower $T_N$ and $T_V$ and higher $T_m$. Fig 13a shows the $T_N$ depression by droplet size reduction. The $T_N$ for a 4.4, 3.4, and 1.4 nm sized droplet drops from 258 K to 242 [106], 220 [106], and 205 [21], respectively. The 1.2 nm sized droplet freezes at 173 °K [107] and the $(H_2O)_{3-18}$ clusters do not form ice even at 120 °K [108].

Molecular undercoordination raises the $T_m$ for a monolayer to 325 K [109] and the skin of bulk water to 310 K [53]. Raman spectroscopy examination of water inside single-walled carbon nanotubes (SWCNT) revealed larger $T_m$ elevation (by as much as 100 °C) [110]. The $T_m$ is bracketed to 105–151

°C for a SWCNT of 1.05 nm across and 87–117 °C for a 1.06 nm SWCNT. Phase changes for the 1.44 and 1.52 nm SWCNTs, transition occur between 15–49 °C and 3–30 °C, respectively. In contrast, the $T_N$ was for the 1.15 nm SWCNT is between −35 and 10 °C.

Fig 13a inset compares the $T_N$ depression of water droplet of slightly more-curved skin deposited on a rough Ag surface. The slightly more curved droplet took 68.4 sec longer to freeze at 269 °K than the contrast deposited on a smooth Ag surface [111]. The formation of the proxy tip due to volume expansion shows completion of frozen. X-ray and neutron reflectometry and AFM [20] revealed that a droplet grown in a certain humidity ambient on the monolayer water film remains "ice-like", hydrophobic, up to $T_m$ at 323 °K (Δm ≥ 0). Evaporation occurs at $T_V$ ~ 338 °K associated with mass loss (Δm < 0). The $T_m$ is higher than 273 °K and the $T_V$ is lower than 373 °K for the droplet compared with the bulk water. These observations verified the QS boundary outward shift, resulting in the supercooling at $T_N$ and $T_V$ and superheating at $T_m$ of the low-dimensional water ice.

Supercooling, also known as undercooling, is the process of lowering the temperature of a liquid below its $T_N$ without it becoming a solid or cooling a gas below $T_N$ without turning into liquid. Once the supercooled water is disturbed with a slapping impulse, it soon becomes "instant ice". Supercooled water occurs in the form of small droplets in clouds and plays a key role in the processing of solar and terrestrial radiative energy fluxes. Supercooled water is also important for life at subfreezing conditions for the cryopreservation of living cells, and for the prevention of hydrate formation in nature gas pipelines. Superheating is the opposite. Once the superheated water is disturbed by adding impurity such as sugar, the superheated water will explode.

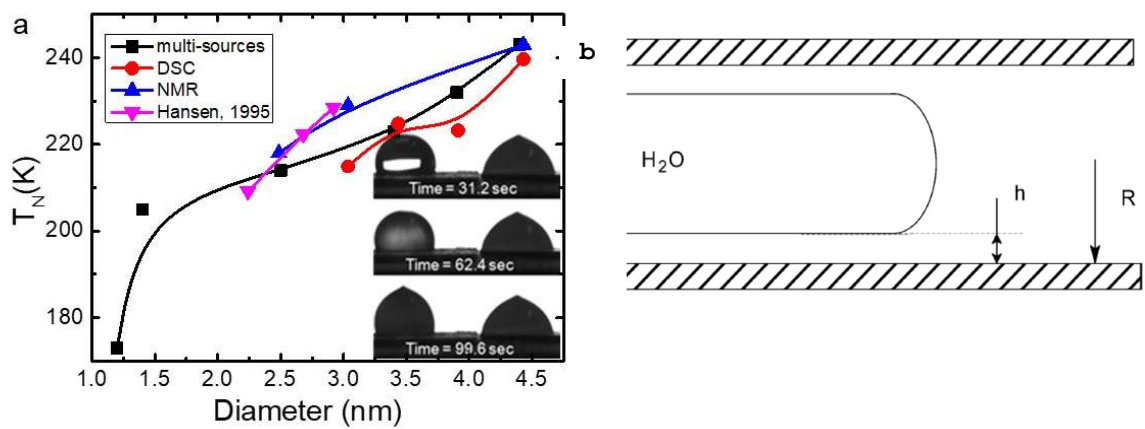

**Fig 13 | Nano-supersolid $T_N$ depression – supercooling** [48, 53]. Skin supersolidity takes the full responsibility for (a) water skin toughness, ice slipperiness, and (b) droplet-size and surface-curvature (inset b) resolved $T_N$ depression (called supercooling) [106, 112]. Droplet of slightly more-curved surface freezes 68.4 sec later at 269 °K than the contrast deposited on smooth Ag surface [111].

## 14. Channel transport: superfluidity and supersolidity

*− Electrostatic repulsivity and mechanical elasticity foster the interface superfluidity* [48]

Superfluidity of water droplet transporting in a hydrophobic channel is of great importance to many subject areas such as cell culturing. An air gap of 0.5–1.0 nm thick exists between the channel wall and the fluid [113] because of the electrostatic repulsion between counter parts at relative motion. For the hydrophobic contact, the water has a 3.8 Å thick skin of 0.71 g/cm$^3$ mass density and there is a 0.6 nm hydrophobic gap between the liquid and the SiO$_2$ support [114]. The presence of a h = 0.6 nm thin vapor (low density) layer separating water and the hydrophobic surface makes the nanoscaled water travelling through the channel much faster than the fluid theory can predict [19], see Fig 13b.

The hydrophobicity, superfluidity, superlubricity, and supersolidity (extended from $^4$He solid) (called 4S) share the same mechanism – elastic, repulsive, and frictionless at motion. Wenzel-Cassie-Baxter's law suggests that nanoscaled roughening makes a hydrophobic surface even more hydrophobic and a hydrophilic surface more hydrophilic.

It is recommended that the fluid skin has a double layer of charge separation without knowing how the double layer is formed [115]. From the BOLS-NEP point of view, the skin H−O contraction and the dual process of oxidation endows the surface with excessive electrons in the form of dipoles. Therefore, the 4S is readily attributed to the high elasticity and the repulsivity of skin dipoles at the contacting interfaces. Particularly, the hydrophobic-hydrophilic cycling transition by an UV excitation or plasma radiation evidences the removal and recovery of the solid surface dipoles. For a solid surface of entrapment dominance, the roughening raises the apex curvature and hence enhances the hydrophobicity or hydrophilicity.

## 15. Perspectives

*− Bonding and electronic dynamics could promisingly lead to the core science of water ice* [89]

To summary, molecular undercoordination matters the performance of low-dimensional water ice by creating a supersolid phase through O:H–O bond cooperative relaxation, O 1$s$ binding electron entrapment, nonbonding lone pair polarization, and specific-heat dispersion. Water prefers the ordered, tetrahedrally-coordinated, fluctuating structure covered with a supersolid skin. O:H–O performs as an asymmetrical, coupled oscillator pair. The cooperativity and the segmental specific-heat disparity of the O:H–O dictate the extraordinary adaptivity, recoverability, sensitivity of water and ice when subjecting to a perturbation.

The O:H–O specific-heat disparity results in the Vapor, Liquid, QS, Ice $I_h$ and $I_c$, and XI phases of mass density oscillation. The QS phase of negative thermal expansion transits the density from maximum to minimum and defines the critical temperatures of melting $T_m$, evaporating $T_V$, and ice nucleation $T_N$. Molecular undercoordination raises the $T_m$ and lowers the $T_V$ and $T_N$ by outward shifting the QS boundary though Einstein's relation $\Theta_{Dx} \propto \omega_x$, which results in the observed supercooling at freezing and evaporating and superheating at melting of the nanometer sized water.

The polarization creates a supersolid skin phase that is hydrophobic, viscoelastic, and frictionless. The supersolid skin of ice and liquid share the common H–O stiffness of 3450 cm$^{-1}$ characteristics, which limits the energy dissipation dynamics of the hydrated electrons and high-frequency H–O phonons. The highly elastic adaptivity of the O:H and the surface electrostatic repulsivity stem the ice slipperiness, water skin toughness, and the superfluidity of water droplet traveling in hydrophobic channels. Surface curvatures enhances the supersolidity and widens the temperature range of supersolidity, which raises the skin reactivity and stabilizes droplet and bubble in mechanical strength and thermal durability of droplets and bubbles. Reproduction of the Mpemba paradox – warm water cooling more quickly, evidences for the skin high thermal diffusivity and lower specific-heat.

Progress demonstrated the efficiency and essentiality of the rules and concepts, particularly, the BOLS-NEP and coupled O:H−O bond theories, in dealing with low-dimensional water ice, which opened the door and paved the path directing to the core chemistry and physics of water and ice. Thinking about water ice and other molecular crystals from the perspective of coupled inter- and intramolecular bonding and electronic dynamics would be even more challenging, fascinating, promising, and rewarding.

## Declaration

No conflicting interest is declared.

## Acknowledgement

Financial support received from the National Natural Science Foundation (No. 21875024) of China and valuable input from collaborators of earlier contributions are gratefully acknowledged.